\documentstyle[12pt]{article}

\def\Z{Z \!\!\! Z}
\def\C{{\mathchoice {\setbox0=\hbox{$\displaystyle\rm C$}\hbox{\hbox
to0pt{\kern0.4\wd0\vrule height0.9\ht0\hss}\box0}}
{\setbox0=\hbox{$\textstyle\rm C$}\hbox{\hbox
to0pt{\kern0.4\wd0\vrule height0.9\ht0\hss}\box0}}
{\setbox0=\hbox{$\scriptstyle\rm C$}\hbox{\hbox
to0pt{\kern0.4\wd0\vrule height0.9\ht0\hss}\box0}}
{\setbox0=\hbox{$\scriptscriptstyle\rm C$}\hbox{\hbox
to0pt{\kern0.4\wd0\vrule height0.9\ht0\hss}\box0}}}}
\def\R{{\mathchoice {\setbox0=\hbox{$\displaystyle\rm R$}\hbox{\hbox
to0pt{\kern0.4\wd0\vrule height0.9\ht0\hss}\box0}}
{\setbox0=\hbox{$\textstyle\rm R$}\hbox{\hbox
to0pt{\kern0.4\wd0\vrule height0.9\ht0\hss}\box0}}
{\setbox0=\hbox{$\scriptstyle\rm R$}\hbox{\hbox
to0pt{\kern0.4\wd0\vrule height0.9\ht0\hss}\box0}}
{\setbox0=\hbox{$\scriptscriptstyle\rm R$}\hbox{\hbox
to0pt{\kern0.4\wd0\vrule height0.9\ht0\hss}\box0}}}}
\def\e0{\epsilon}
\def\e{\varepsilon}
\def\bc{\begin{center}}
\def\ec{\end{center}}
\def \be{\begin{equation}}
\def \eq{\end{equation}}
\def \bee{\begin{eqnarray}}
\def \eqq{\end{eqnarray}}

\begin{document}

\begin{flushright}
{hep-th/0102052}
\end{flushright}

\title{GRADED CONTRACTIONS OF VIRASORO ALGEBRAS}
\author{I.V.Kostyakov, N.A.Gromov, V.V.Kuratov \\
Department of Mathematics, \\ 
Syktyvkar Branch of IMM UrD RAS, Russia \\
e-mail: kuratov@dm.komisc.ru}
\maketitle

\begin{abstract}
{We describe graded contractions of Virasoro algebra.
The highest weight representations of Virasoro algebra are constructed.
The reducibility of representations is analysed.
In contrast to standart representations the contracted ones
are reducible except some special cases.
Moreover we find an exotic module with null-plane on fifth level.
}
\end{abstract}

\section{Introduction}
The Virasoro algebra takes an important part in strings theory,
integrable models and two-dimensional conformal field theory.
Resently it has been investigated the extensions 
of Virasoro algebra leading to $W$-algebras, which contain the 
Virasoro algebra as a subalgebra.
$W$-algebras are appeared in integrable models~\cite{1,2}
and then in two-dimensional conformal field theory~\cite{3}.  
Thereafter the notions of $W$-strings, $W$-gravitation have arisen
and $W$-models of conformal field theory have been created~\cite{4}.
Hence the extensions of symmetries on the base of the Virasoro
algebra are very fruitful.

A different way of extension of Virasoro algebra is a contraction.
In this case new nonsemisimple algebras
containing the Virasoro algebra as a
subalgebra are arisen.
 Contractions of affine Kac-Moody algebras
were studied in \cite{5,6} and similar effect was obtained \cite{7}.
Nonsemisimple algebras are used as algebras of internal
symmetries of different physical models:  gauge theories~\cite{8},
 WZNW models~\cite{9,10}. Sugavara construction was developed for
nonsemisimple algebras~\cite{11}.
The purpose of this paper is the investigation of contractions
of Virasoro algebra and its representations with the help of the
graded contraction method~\cite{7,12}.

This paper is organized as follows.
In the second section  we briefly present  the main ideas
of graded contraction method. In Sec.3 the grading
of Virasoro algebra by cyclic groups $\Z_2$ is suggested.
In Sec.4 we recall the base facts of reducible theory of modules
over Virasoro algebra and  describe their $\Z_2$-grading.
Sec.5 contains the results of some contractions
of Virasoro algebra. In Sec.6  contractions of modules
and some problems of their reducibility are investigated.

\section{Graded contractions of Lie algebras and their representations}
A Lie groups (algebras) contraction makes possible to get new Lie 
groups (algebras) from some initial ones. Moreover one can construct
a representation of contracted algebras from a representation of 
starting algebra. The graded contractions \cite{6,12} are defined in such a 
way that the grading of both Lie algebra and its representation are preserved. 

First we  recall some definitions.
Let a Lie algebra $L$ is graduated by an abelian group
$G$ as follows
\begin{equation}
 L=\bigoplus L_j ,\ \
  [L_j,L_i] \subseteq L_{j+i}, \ \ j,i \in  G.
\label{1}
\end{equation}
A Lie algebra $L^{\varepsilon}$ is called a $G$-graded contraction  
of the algebra $L$ if $L^{\varepsilon}$ is isomorphic to $L$ as vector space,
$L^{\varepsilon}$ has grading~(\ref{1}) and new commutation relations are
\begin{equation}
[L^{\e}_j,L^{\e}_k]_{\e}:=\e_{jk}[L_j,L_k] \subseteq \e_{jk}L^{\e}_{j+k},
\label{2}
\end{equation}
where the matrix $\e$ is a solution of the equations
\begin{equation}
\e_{jk}\e_{m,j+k}=\e_{km}\e_{j,m+k}=\e_{mj}\e_{k,m+j},
\ \ \e_{jk}=\e_{kj}.
\label{3}
\end{equation}
The first set of equations in~(\ref{3}) is the consequence of Jacoby identity
and the second one follows from antisimmetry of commutator.

$G$-grading of $L$-module $V$ is obtained from decomposition~(\ref{1})
and looks as follows
\begin{equation}
V=\bigoplus V_m ,\ \ L_jV_m \subseteq V_{j+m},
\label{4}
\end{equation}
\begin{equation}
[L_j,L_k]V_m=L_jL_kV_m-L_kL_jV_m \subseteq V_{j+k+m}.
\end{equation}
A $G$-graded contraction of algebra representation is defined by
\begin{equation}
L_j^{\e}V_m^{\e,\psi}:=\psi_{jm}L_jV_m \subseteq \psi_{jm} V^{\e,\psi}_{j+m}.
\label{6}
\end{equation}
where the matrix $\psi$ satisfy the following equations~\cite{7}
\begin{equation}
\e_{jk}\psi_{j+k,m}=\psi_{km}\psi_{j,k+m}=\psi_{jm}\psi_{k,j+m}.
\label{7}
\end{equation}
Hereafter we shall use for grading only cyclic group $G=\Z_2$.
It is convenient to write down the vector space $V$ and 
action of algebra  $L$ for $\Z_2$-grading in the  form
\begin{equation}
V={V_0\choose V_1}, \ \ L V={L_0 \ L_1 \choose L_1 \ L_0}{V_0 \choose V_1}=
{L_0V_0+L_1V_1 \choose L_1V_0+L_0V_1},
\end{equation}
explicitly showing the structure of grading.
Then we have for contracted module 
\begin{equation}
 L^{\psi} V={L_0 \ L_1 \choose L_1 \ L_0}^{\psi} {V_0 \choose V_1}=
{\psi_{00}L_0 \ \psi_{11}L_1 \choose \psi_{10}L_1 \ \psi_{01}L_0}
{V_0 \choose V_1}.
\label{9}
\end{equation}

\section{Grading of Virasoro algebra}
The Virasoro algebra $\cal L$(c) is the central extension of 
the algebra of vector fields on the circle with the basis 
$c,l_n \ (n \in \Z)$ and the commutation relations
\begin{equation}
[l_n,l_m]=(n-m)l_{n+m}+{c \over 12}(n^3-n)\delta_{n+m,0},
\label{Vir}
\end{equation}
where $c$ is the central charge commuting with all generators.
This algebra has the natural $\Z$-grading:
deg $l_k=k $, deg $c=0 $
and may be graded also by cyclic groups $\Z_n$.
The method of $\Z_2$-grading of the Virasoro algebra consists in follows.
The set of the Virasoro algebra generators is divided on even 
$ L_0=\{A_n, \, c\}$ and odd $L_1=\{B_n\}$ parts,
where
\begin{equation}
A_n={1 \over 2}\left(l_{2n}+{c\over 8}\delta_{n,0}\right),
\quad
B_n={1 \over 2}l_{2n+1},
\label{A=B=}
\end{equation}
then the $Z_2$-grading conditions are held
$$
{\cal L} = L_0 \bigoplus L_1,\ \
[ L_0,L_0] \subseteq L_{0},\ \
[ L_0,L_1] \subseteq L_{1},\ \
[ L_1,L_1] \subseteq L_{0}.
$$
and the commutation relations of Virasoro algebra for new generators look as 
\begin{eqnarray}
[A_n,A_m] & = & (n-m)A_{n+m}+{2c \over 12}(n^3-n)\delta_{n+m,0}, \cr
[B_n,B_m] & = & (n-m)A_{n+m+1}+
{2c \over 12}(n-{1\over2})(n+{1\over2})(n+{3\over2})\delta_{n+m+1,0}, \cr
[A_n,B_m] & = & (n-m-{1\over 2})B_{n+m} .
\label{NVir}
\end{eqnarray}
The first line of equations means that the subalgebra $\{A_n,c \}$ is
again Virasoro algebra but with central charge 2c.
In the case of $\Z_n-$grading  we have Virasoro
subalgebra with  central charge $nc$. This looks like the result 
of generating the orbifold algebras by $\Z_{\lambda}$-orbifold
induction procedure $c\rightarrow \lambda c$~\cite{Bor}.

\section{Reducibility of representations. $\Z_2$-grading}
In this section we recall a basic notions and results
of the theory of Virasoro algebra representations \cite{13} and 
describe their $\Z_2$-grading. 
Let  $M$ be the module over ${\cal L}(c)$ and there is the vector
$ | v \rangle$ such that
\begin{equation} 
l_0| v \rangle=h| v \rangle , \quad
 \hat c | v \rangle=c| v \rangle, \quad
 l_k| v \rangle=0, \ k>0,
\label{condition}
\end{equation}
then $| v\rangle$ is a highest vector.
The space spanned by vectors $l_{-i_1}\ldots l_{-i_m}| v  \rangle,$
where $i_1 \geq \ldots \geq i_m > 0 $, 
form a highest vector representation $M(h,c)$.
All vectors of module are classified by levels.
The level number is given by $ n=i_1+\ldots +i_m$.
For example $l_{-2}l_{-1}|~v~\rangle$ belongs to the third level.
Therefore the module has natural $\Z$-grading $M(h,c)=\bigoplus M^n(h,c)$.
The number of basis vectors on the $n$-th level is defined by the number
$p(n)$ of partitions of $n$ on positive integer numbers: dim$M^n(h,c)=p(n)$. 
Some values of $p(n)$ are given below:
$$
p(0)=1,\ p(1)=1,\ p(2)=2,\ p(3)=3,\ p(4)=5,\ p(5)=7.
$$

The investigation of reducibility conditions of module is an important
problem. A module is called degenerate if there is a null-vector
$| \chi_n \rangle$, such that
$$
l_0| \chi_n \rangle=(h+n)| \chi_n \rangle ,
\quad l_k| \chi_n \rangle=0, \quad  k > 0,
$$
where $n$ is called the degeneracy of level. 
The degenerated module contains the submodule with highest vectors
$| \chi_n \rangle$ on the $n$-th level and therefore is a reducible
module. Null-vector is found as a linear combination of basis vectors with
unknown coefficients. The conditions of null-vectors lead to the system
of linear equations with two parameters $h$ and $c$.
The number of equations  and the number of unknowns are given by 
$
N_1(n)=p(n-1)+p(n-2)
$
and
$
N_2(n)=p(n)-1
$
respectively.

On the second level we have two equations and only one unknown.
The relation between h and c on this level is given by
$$
h={5-c\pm\sqrt{(c-25)(c-1)}\over16}.
$$
The number of equations on the next levels increases faster then
the number of unknowns and therefore the existence of null-vectors
becomes a rarity. The problem of searching of null-vectors is 
completely solved with the help of bilinear form on the $M(h,c)$.
Let  $w$ be an antiautomorphizm ${\cal L}(c)$ defined by
$w(l_i)=l_{-i}$, $ w(c)=c,$
then there is a symmetrical bilinear form $(\ \mid \ )$, such that
$(a| l b)=(w(l)a| b),$
where $a,b \in M(h,c),\ l \in {\cal L}(c)$.
Let the  $p(n) \times p(n)$ matrix $K^n(h,c)$ is the matrix of 
bilinear form of basis vectors on $n$-th level. The vanishing of 
determinant of matrix $K^n(h,c)$ is the 
condition of existence of null-vector~\cite{13}.
There are some possibilities depending on the values of parameters
$h,c$: 1) the module is irreducible, then there are no null-vectors;
2) the submodules generated by null-vectors are embedded one into another;
3) there are two submodules containing all other submodules.
In the last case the modules $M(h,c)$ are defined by
$$
c=1-{6(q-p)^2 \over pq },
\qquad h={(qr-ps)^2-(q-p)^2 \over 4pq},
$$
where $0<r<p,\ 0<s<q$, for integers $r,s$ and $q=p+1=3,4,\ldots$.
The last modules  describe the space of fields of minimal models
in conformal fields theory. Degeneracy of module leads to the differential
equations on the fields correlators \cite{BPZ}.

We are coming now to the describing of $\Z_2$-grading of module $M(h,c)$.
We shall consider that the Virasoro algebra is $\Z_2$-graded by the above 
described method. The representation space may be divided on two subspaces
$$
M(h,c)=M_0(h,c)+M_1(h,c),
$$
$$
M_0(h,c)=\bigoplus M^{2n}(h,c), \quad
M_1(h,c)=\bigoplus M^{2n+1}(h,c).
$$
The grading of elements $l_{-i_1}\ldots l_{-i_m}$
of enveloping algebra is realized by decomposition on two subset
with odd and even value of sum $i_1+\ldots +i_m$.
The vectors from $M_0(h,c)$ and $M_1(h,c)$ in terms of generators 
$A$ and $B$ look as
\begin{equation}
AA \ldots AA \underbrace{BB \ldots BB}_{k} | v \rangle ,
\label{form}
\end{equation}
where $k$ is even for $M_0(h,c)$ and odd for $M_1(h,c)$.
Then the following equations of $\Z_2$-grading are realized
$$
AM_0 \subseteq M_0,\ AM_1 \subseteq M_1,\
BM_0 \subseteq M_1,\ BM_1 \subseteq M_0.
$$
The structure of the $\Z_2$-grading module may be written as
\begin{equation}
L M={A \ B \choose B \ A}{M_0 \choose M_1}=
{A M_0+B M_1 \choose B M_0+A M_1}.
\label{LM}
\end{equation}
Let us note that antiautomorphism $w$ in $\Z_2$-graded notation
looks as 

$
w(A_i)=A_{-i}, \ w(B)=B_{-i+1}, \ w(c)=c.
$

\section{$\Z_2$-graded contraction of Virasoro algebra}
Consider the contraction given by solution 
$
\e^{\alpha}=\left(
\begin{array}{cc}
1&1\\
1&0\\
\end{array}
\right)
$ 
of equations~(\ref{3}).
The commutation relations (\ref{NVir}) take the form
\begin{eqnarray*}
[A_n,A_m] & = & (n-m)A_{n+m}+{2c \over 12}(n^3-n)\delta_{n+m,0}, \cr
[A_n,B_m] & = & (n-m-{1\over 2})B_{n+m}, \qquad
[B_n,B_m]  =  0.
\label{CVir}
\end{eqnarray*}
The contracted algebra has the structure of {\it semidirect}
sum of Virasoro algebra with double central charge ${\cal L}(2c)=\{A_n, 2c\}$
and infinite dimensional abelian one $\{B_n\}$.
It may be called as inhomogeneous Virasoro algebra.
The second solution of equations~(\ref{3})
$
\e^{\beta}=\left(
\begin{array}{cc}
1&0\\
0&0\\
\end{array}
\right)
$
leads to the contracted Virasoro algebra, which is a {\it direct} sum
of ${\cal L}(2c)$ and abelian $\{B_n\}$
\begin{eqnarray*}
[A_n,A_m] & = & (n-m)A_{n+m}+{2c \over 12}n(n^2-1)\delta_{n+m,0}, \cr
[A_n,B_m] & = & 0, \qquad \quad [B_n,B_m]  =  0.
\end{eqnarray*}

\section{$\Z_2$-Grading contractions of Virasoro modules}

It is easily to understand the structure of contracted $\Z_2$-module 
using the relations (\ref{LM}) and (\ref{9}).
We take the solution of equations~(\ref{7}) 
$
\psi^{\alpha}=\left(
\begin{array}{cc}
1&1\\
1&0\\
\end{array}
\right).
$
In this case
\begin{equation}
L M={A \ B \choose B \ A}{M_0 \choose M_1}=
{A M_0 \ \quad \choose B M_0+A M_1}.
\end{equation}
Therefore all properties of contracted module may be obtained
by assuming
\begin{equation}
B M_1=0.
\label{BM1}
\end{equation}
which means that all vectors of the form (\ref{form}) with $k \geq 2$
are vanish. In other words the contracted module is obtained from the initial
one by  factorization with respect to condition $ B B | v \rangle =0$.

It follows from (\ref{A=B=}) that the parameters of contracted
$M_{\psi}(h,c)$ and initial $M(h_0,c_0)$ modules are
connected by relations
$
c=2c_0,\ h={1\over 2}(h_0+{c_0\over 8}).
$
It is easily to understand that the structure of vectors on the even level $2n$
of contracted module and the structure of vectors on the level $n$ of initial
one are identical.
Therefore
dim$M_{\psi}^{2n}(h,c)= {\rm dim} M^{n}(h,c)=p(n)$.
The vectors of odd level look as
$A_{-i_1} \ldots A_{-i_n}B_{-q} | v \rangle$.
The dimension of odd level is given by
\begin{equation}
{\rm dim}M_{\psi}^{2n+1}(h,c)=\sum_{k=0}^{n}p(k).
\label{dim}
\end{equation}

Phisical requirement that the spectrum must be bounded from below
leads to the highest vector condition (\ref{condition}). Since
any element $l_k$ is expressed as multiple commutators of
$l_1$ and $l_2$, then in (14) instead of $l_k| v \rangle=0,\ k>0$
it is enough to impose two conditions:
$l_{1}| v \rangle=0,\ l_{2}| v \rangle=0$.
Then the highest vector conditions~(\ref{condition}) are simplified
\begin{equation}
A_0 |v \rangle=h | v \rangle , \ \hat c | v \rangle=c | v \rangle,\
B_0 | v \rangle=0, \ A_1 | v \rangle=0.
\label{condition1}
\end{equation}
For $\e^{\alpha}$-contracted Virasoro algebra these conditions 
needs to be modified by adding the equation
$
A_2 | v \rangle=0,
$
since otherwise the generators $A_k,\ k>1$ are not generated
by commutators of $B_0$ and $A_1$.
Thus the existence conditions of null-vector on the level $n$ are
\begin{equation}
A_0| \chi  \rangle=(h+n)| \chi  \rangle,\
B_0| \chi  \rangle=0, \ A_1| \chi  \rangle=0,\
A_2 | \chi  \rangle=0.
\label{condition2}
\end{equation}
The increase of number of equations on even levels with compared
to~(\ref{condition1}) leads to vanishing of the null-vectors on these levels.
It is possible to verify this fact by direct calculations.
The relations $B_0| \chi  \rangle=0$
for odd level  are held due to~(\ref{BM1}).
Thus the number of conditions remains the same
therefore one would expect the presence of null-vectors on odd levels.
The number $N_1$ of equations and the number $N_2$ of unknown 
coefficients, which  defined the existence of the null-vectors
on $(2n+1)$-th level are easily obtained from~(\ref{dim}) and
(\ref{condition2}) 
$$
N_1= \sum_{k=0}^{n-1} p(k) + \sum_{k=0}^{n-2} p(k),
\qquad \quad
N_2=\sum_{k=0}^{n} p(k)-1
$$
It follows from this equations that $N_1=N_2$ up to eleventh level
and $N_1>N_2$ for higher-order levels.

Let us investigate some first levels in details.
One find by direct calculations that there are null-vectors on the first and 
on the third levels

$
| \chi \rangle_1 = B_{-1} | v \rangle,
\qquad
| \chi \rangle_3 = \left ( B_{-2} - {5 \over 4h+2} A_{-1}B_{-1} 
\right ) | v \rangle,
$

\noindent
for any values of $h,c$. On the fifth level the null-vector looks as

$ | \chi  \rangle_5 = \left ( B_{-3}+\alpha A_{-2}B_{-1}+
\beta A_{-1}B_{-2}+\gamma A_{-1}^2B_{-1} \right ) | v \rangle,
$

\noindent
where $\alpha, \beta, \gamma$ are solutions of system 
with parameters $h$ and $c$

$
\hspace{25mm}  (4h+6)\beta \qquad \qquad \qquad  =7,
$

$
\hspace{6mm} (4+8h+c)\alpha + 15\beta +(12h+6)\gamma = -9,
$

$
\hspace{26mm} 6\alpha   + \  5\beta + \ (8h+8)\gamma  =0.
$

\noindent
It is immediately follows from the first equation that
there are no null-vectors on the straight line $h=-3/2$. The
substitution of the first equation into the others reduces the analysis
of system to possible disposition of a straight lines on the plane 
$(\alpha, \gamma)$. Let $D(h,c)$ be the determinant of matrix 
composed from the coefficients of the left part of system
and let $D_i(h,c)$,  $i=1,2$ be the determinant of the matrix, which is 
obtained from the previous one by the replacement of $i$-th column with
the column of the right part of the system.
There are three possibility depending on the parameters $h$ and $c$:
1) the straight lines are crossed -- in this case there are null-vectors;
2) the straight lines are parallel -- in this case there are not null-vectors;
3) the straight lines are coincided -- there are null-vectors,
moreover the null-vectors become two-dimensional or "null-plane".
The second case is realized on the straight line
$h=-{3\over2},$
and two curves $D(h,c)=0$ on (h,c)-plane, i.e.
$$
h={-3-c\pm\sqrt{(c-25)(c-1)}\over16}, 
$$
The third case is a possibile at two points on (h,c)-plane: 
$A(h_1=-{3\over2}, c_1=26)$ and $B( h_2={11\over24}, c_2=-{184\over105})$
which are the  intersection of the curve $D(h,c)=0$
and the straight line $D_2(h,c)=0$, i.e. $c=(176-496h)/35.$
However the point $A$ is on the straight line $h=-{3\over 2}$, where there 
are not null-vectors. So the null-plane on the fifth level appears at 
the unique values of parameters corresponding to the point 
$B\left(h_2,c_2\right)$.
Similar result was obtained in paper~\cite{15} where it was
shown that certain Verma modules over the $N=1$ 
Ramond algebra contain degenerate two-dimensional singular vector spaces.
For the rest values of parameters $h$ and $c$ the first case is realized.

On the seventh and ninth levels the numbers of equations and
unknown coefficients are coincided and  are equal to 6 and 11
respectively. However the calculations become too cumbersome,
although it is clear that null-vectors are absent only in a special cases.
Starting with eleventh level the number of equations begins to exceed
the number of unknowns $ N_1(11)=19,\ N_2(11)=18  $ and the 
situation becomes apparently similar to the case of noncontracted module.
The attempt of using the ordinary bilinear form as in the case of 
noncontracted algebra does not leads to the solution of the problem of
searching of null-vectors. The point is that all vectors on odd levels
are orthogonal  to each other due to~(\ref{BM1}). It is an ordinary thing 
for a finite Lie algebra that the Killing form is degenerated under contraction. 
At this point it may be useful the results of the work~\cite{Mont}, where the 
way of construction of nondegenerate forms for the contracted algebras was
proposed.

\end{document}